\documentclass[a4paper,aps,prl,floatfix,twocolumn, superscriptaddress,nobibnotes]{revtex4-2}

\usepackage[pdftex]{graphicx}
\setkeys{Gin}{width=0.9\columnwidth}

\usepackage{times}
\usepackage{verbatim}
\usepackage{bbold}
\usepackage{bbm}
\usepackage{lipsum}

\usepackage{amsmath}
\usepackage{amsfonts}
\usepackage{color}
\usepackage[english]{babel}
\usepackage{microtype}
\usepackage{comment}
\usepackage[normalem]{ulem}
\usepackage{float}
\usepackage{physics}

\usepackage{soul,xcolor}
\setstcolor{red}

\usepackage{relsize}

\usepackage[hidelinks,unicode=true]{hyperref}
\hypersetup{
colorlinks=true,
linkcolor=blue,
urlcolor=blue,
citecolor=blue,
pdfhighlight=/N}

\usepackage{comment}

\usepackage{relsize}

\newcommand{\KD}[1]{\textcolor{purple}{[KD: #1]}}

\graphicspath{{figs/}}

\begin{document}

\title{Cross-order induced behaviors in contagion dynamics on higher-order networks}


\author{Kaloyan Danovski}
\affiliation{Department of Engineering, Universitat Pompeu Fabra, 08018 Barcelona, Spain}

\author{Sandro Meloni}
\affiliation{Institute for Cross-Disciplinary Physics and Complex Systems (IFISC, CSIC-UIB), 07122 Palma de Mallorca, Spain}
\affiliation{Institute for Applied Mathematics ``Mauro Picone'' (IAC) CNR, Via dei Taurini 19, 00185 - Rome, Italy}
\affiliation{``Enrico Fermi'' Research Center (CREF), Via Panisperna 89A, 00184 - Rome, Italy}

\author{Michele Starnini}
\email{Corresponding author: michele.starnini@upf.edu}
\affiliation{Department of Engineering, Universitat Pompeu Fabra, 08018 Barcelona, Spain}

\date{\today}

\begin{abstract}
Recent studies have shown that novel collective behaviors emerge in complex systems due to higher-order interactions. 
However, the way in which the structural correlations of these interactions shape such behaviors remains a significant gap in current research. 
To address this, we use signatures of higher-order behaviors (HOBs) to identify the underlying dynamical rules, or higher-order mechanisms (HOMs). 
In this work, we compare several HOB measures derived from information theory. 
Utilizing a simplicial SIS contagion model, we demonstrate that simpler, computationally efficient measures can serve as robust indicators of HOMs
. 
We uncover 
the novel phenomenon of cross-order induced behaviors, where behavioral signatures emerge at interaction orders where no direct mechanism is present. 
Crucially, these cross-order HOBs are not simply induced by structural correlations ---such as nestedness and hyperedge overlap--- but they appear in the neighborhood of any HOM.
Among the information-theoretic measures we tested, synergy is the most reliable indicator of the true order where the underlying mechanism is at play. 
These findings offer new insights into the relationship between the network structure and observed dynamics of higher-order systems.
\end{abstract}

\maketitle

\section{Introduction}

Many of the systems of scientific and general interest today are complex systems, made up of many simple units interacting in non-simple ways. 
In many of these systems, the set of interactions are naturally described by a network: a mathematical object whose structure has repeatedly shown to be crucial for modelling the system's dynamics~\cite{boccaletti2006complex}. 
However, recently extensions of the classical network framework to include so-called higher-order interactions---e.g. interactions involving three or more nodes at the same time---have been proposed~\cite{battistonNetworksPairwiseInteractions2020,Musciotto2021detecting,Lotito2022detecting,Majhi2022review,Bick2023review,Boccaletti2023review,Ceria2025hoi}. 
Such interactions are essential for representing real-world processes which cannot be decomposed into a set of pairwise interactions~\cite{Neuhauser2020reducibility,meloni2026reducibility,tan2026reducibility,llabres2026reducibility}, such as group social interactions~\cite{benson2018simplicial,Iacopini2022social, bettiDyadHigherorderStructure2025,Arregui2024social,PerezMartinez2025social}, ecological competition~\cite{levine2017beyond}, and neural processing units~\cite{petri2014homological, giusti2016two, santoro2022unveiling}, to name a few.

Moreover, recent research showed that structural correlations between higher-order structures are a fundamental determinant of how these dynamical processes unfold. 
In particular, inter-order hyperedge overlap which refers to a hierarchical correlation structure where smaller hyperedges are entirely contained within larger ones, is a common feature of real-world social systems. 
These structural correlations significantly alter transitions in dynamical processes~\cite{Lee2021overlap,Larock2023overlap,kimContagionDynamicsHypergraphs2023,Zhang2024synch,Leon2024synch,Gallo2022synch,Gambuzza2021synch}. 
For instance, 
the emergence of explosivity and bistability in synchronization dynamics often requires specific levels of hyperedge overlap~\cite{maliziaHyperedgeOverlapDrives2025,Lamata2025overlap,Moriame2025synch,Muolo2025synch,Zhang2023synch}. 
In the context of contagion dynamics, the degree of overlap can dualistically lower invasion thresholds while reducing the size of outbreaks~\cite{burgioTriadicApproximationReveals2024,malizia2025overlap}. 
Beyond contagion, nested structures are found to promote the emergence of cooperation in evolutionary dynamics when they comprise rich, uniform nested dyadic structures~\cite{xuNestedStructuresHigherorder2024}.


The interplay between these structural constraints and the resulting dynamics necessitates a clear distinction between higher-order mechanisms (HOMs) and higher-order behaviors (HOBs)~\cite{rosasDisentanglingHighorderMechanisms2022}. 
In this context, \textit{mechanisms} refers to the (potentially dynamical) rules of interaction between system units, while \textit{behaviors} refers to the statistical patterns in the (potentially time-dependent) state of the variables that describe the system. 
Both are considered higher-order when they involve more than 2 variables in a way that is irreducible to purely pairwise relationships. 
Understanding the relationship between these two concepts is vital for network inference; ideally, one could identify the presence of complex hidden mechanisms by measuring their behavioral signatures. 
Yet, HOMs and HOBs do not always correspond directly. 
There are instances where HOMs leave no detectable behavioral mark, or conversely, where purely pairwise interactions generate behaviors that appear higher-order~\cite{robiglioSynergisticSignaturesGroup2025,rosasDisentanglingHighorderMechanisms2022,Neuhauser2020reducibility,meloni2026reducibility,tan2026reducibility,llabres2026reducibility}.

In this work, we compare several measures of HOBs based on the Partial Information Decomposition \cite{williamsNonnegativeDecompositionMultivariate2010} to identify which serves as the most reliable indicator of underlying HOMs, while assessing the robustness of these signals as the interaction order increases.
Utilizing a simplicial SIS contagion model, we explore how these signals emerge and scale across both simplicial complexes and hypergraphs. 
We specifically investigate the phenomenon of cross-order induced behaviors, where mechanisms at one order trigger behavioral signatures at another. 
Our findings suggest that simpler, computationally efficient measures can often capture the same qualitative behaviors as more complex ones, provided they are properly conditioned to isolate dynamical dependence. 
Furthermore, these induced behaviors emerge when no structural correlations, such as nestedness or hyperedge overlap, are present, offering new insights into the interplay between network structure and observed higher-order dynamics.

\section{Results}
\label{sec:res}

\subsection{Measures of higher-order behavior}
\label{ssec:res-measures}


We consider the simplicial SIS model \cite{iacopiniSimplicialModelsSocial2019}, as one of the simplest stochastic systems for generating higher-order behaviors. 
For a simplicial complex of maximum order $L=2$, the simplicial SIS model is defined by a pairwise infectivity $\lambda_1$  (the classic rate of individual, pairwise infection) and a higher-order infectivity $\lambda_2$ (an additional infection rate for each 2-simplex with one susceptible and two infected nodes).
See Section ``Numerical simulations of the contagion dynamics" in SM for details.
We interpret $\lambda_1$ and $\lambda_2$ as the strength of the pairwise (PWM) and higher-order (HOM) mechanisms, respectively. 
We study the behavior of this model via stochastic simulations on a quenched random simplicial complex (see Section ``Generating random higher-order networks" in SM). 
In all cases, we fix the number of nodes to $N=200$ and the average pairwise degree to $\langle k_1\rangle=20$. We adjust the average degree at orders larger than 1 in each of the experiments described below, in order to provide a sufficient number of both simpleces and purely pairwise cliques for the statistical analysis. 
We also fix the independent recovery rate to $\mu=0.8$ and the initial density of infected nodes to $\rho_0=0.3$.

To quantify HOBs within groups of nodes, we compare different measures from information theory. 
We start by looking at the sum of pairwise transfer entropies 
\begin{equation}
    \mathcal{T}(\vb{X})=\sum_{i,j\neq i}T(X_j\to X_i)=\sum_{i,j\neq i}I\qty(X_i^{(t)};X_j^{(t-1)}|X_i^{(t-1)})
\end{equation} 
as a purely lower-order measure, since it considers only the pairwise mutual information $I(Y;X)$, which does not depend on the joint state of more than two variables. 
The superscripts $(t)$ and $(t-1)$ indicate the state at the current and previous time step, respectively, such that the transfer entropy $T$ captures the information flowing from one source $X_j^{(t-1)}$ to one target variable $X_i^{(t)}$ over one time step, conditioned on the history of the target $X_i^{(t-1)}$.
We use $\mathcal{T}$ as a baseline for comparison with other measures, as it has a much weaker dependence on higher-order mechanisms  \cite{robiglioSynergisticSignaturesGroup2025}.

The simplest way to include information from more than two variables is to use a joint distribution inside the mutual information $I(Y;\vb{X})$, where $\vb{X}=\{X_i\}_{i=1..n}$ is the set of $n$ sources. 
Averaging this quantity over all nodes in the group as targets, we obtain the average mutual information
\begin{equation}
    \mathcal{I}(\vb{X})=\frac{1}{n}\sum_{i=1}^n T(\vb{X}_{-i}\to X_i),
\end{equation}
where $\vb X_{-i}=\vb X \setminus \{X_i\}$.
Since we apply a time delay and condition on the target's history, the average mutual information becomes the average multivariate transfer entropy, but we choose the notation for $\mathcal{I}$ to avoid ambiguity with its pairwise counterpart $\mathcal{T}$.

However, the quantity $I(Y;\vb{X})$ of information that a set of sources provides about a target aggregates many different types of dependence. 
Notably, some amount of information can be stored redundantly in each source variable, while some amount is contained within the group of variables synergistically, i.e., accessible only through the joint state of all variables, not the state of any lower-order subgroups. 
These quantities are called redundancy and synergy, respectively.
Synergistic information is one of the defining features of higher-order behavior, and is analogous to the irreducibility of higher-order mechanisms to lower-order ones (e.g., pairwise interactions).
Furthermore, groups involved in HOMs are expected to contain a greater amount of synergy than those involved in PWMs only, since the dynamical rules of HOMs depend on multiple variables at the same time. 

To measure this effect, we can use the average synergy
\begin{equation}
    \mathcal{S}(\vb{X})=\frac{1}{n}\sum_{i=1}^n \qty[ T\qty(\vb{X}_{-i}\to X_i) - \max_j{T\qty(\vb{X}_{-ij}\to X_i)} ],
\end{equation}
which captures the average amount of information about each variable that is available only when looking at all other variables together.
In the definitions of both $\mathcal{I}$ and $\mathcal{S}$, the conditioning on the target's history ensures we exclude the effects of common drivers between source and target nodes, thus isolating the dynamical dependence between them. 
To test how sensitive our results are to this choice, we also consider the time-lagged but non-conditioned average mutual information $\hat{\mathcal{I}}$ and average synergy $\hat{\mathcal{S}}$
(see Eqs. (1) and (6) in SM Section ``Information-theoretic measures: definition").

While we would expect that an increase in synergy is related to the presence of HOMs, it could also result from an increase in the overall mutual information, even for PWMs alone. 
Thus, it is not clear whether the identifying feature of HOMs is a larger quantity of synergy or instead a dominance of synergy over redundancy, which is reflected in the difference between the two types of information. 
To test which of these observables is a stronger signal of HOMs, we employ the average dynamical O-information \cite{rosasQuantifying2019,stramagliaQuantifyingDynamicalHighOrder2021}
\begin{align*}
    d\Omega(\vb{X}) &\equiv \langle d\Omega(X_i;\vb{X}_{-i})\rangle_i \\
    &= \frac{1}{n}\sum_{i=1}^n \qty[ (1-n)T\qty(\vb{X}_{-i}\to X_i) + \sum_{j=1}^nT\qty(\vb{X}_{-ij}\to X_i) ].
\end{align*}
This quantity captures a balance between various kinds of synergistic and redundant information in the system, such that $d\Omega<0$ indicates a dominance of synergy and $d\Omega>0$ that of redundancy. 
For a group of 3 variables (two sources $\vb{X}$ and one target $Y$), it can be shown that $d\Omega(Y;\vb{X})=R(Y;\vb{X})-S(Y;\vb{X})$, where $R$ and $S$ denote the single-target redundancy and synergy, respectively (see SM Section ``Multivariate information and its decomposition" for details).
We want to test how these measures are affected by the presence and strength of different HOMs. 
To this aim, we compare the value obtained for a given measure $M \in \{ d\Omega, \mathcal{S}, \hat{\mathcal{S}}, \mathcal{I}, \hat{\mathcal{I}}, \mathcal{T} \}$ when applied to groups of nodes interacting only via pairwise mechanisms against those where nodes are additionally involved in higher-order interactions. 
That is, we apply $M$ to the time-series of two distintic sets of 3 nodes, connected either in a 3-clique or in a 2-simplex, obtaining two distributions $P^c_M$ and $P^s_M$, respectively.
Despite being measured on pairwise-connected nodes, $P^c_M$ might be different from zero \cite{robiglioSynergisticSignaturesGroup2025} (see Supplementary Figure 1).
This is partly a result of performing numerical approximations of the information, but mostly because higher-order behaviors are not inherently excluded from pairwise systems, and indeed are observed in many cases, such as the model we study here.
Therefore, we take the statistical distance between the distributions $\delta(P_M)\equiv\delta(P_M^c,P_M^s)=\frac{1}{2}\sum_x|P_M^c(x)-P_M^s(x)|$, which can better quantify the HOB due to the presence of HOMs (e.g., 2-simplices).
See Section ``Information-theoretic measures: measurement" in SM for more details on the estimation procedure and distance calculation.

\begin{figure*}[tbp]
    \centering
    \includegraphics[width=0.8\linewidth]{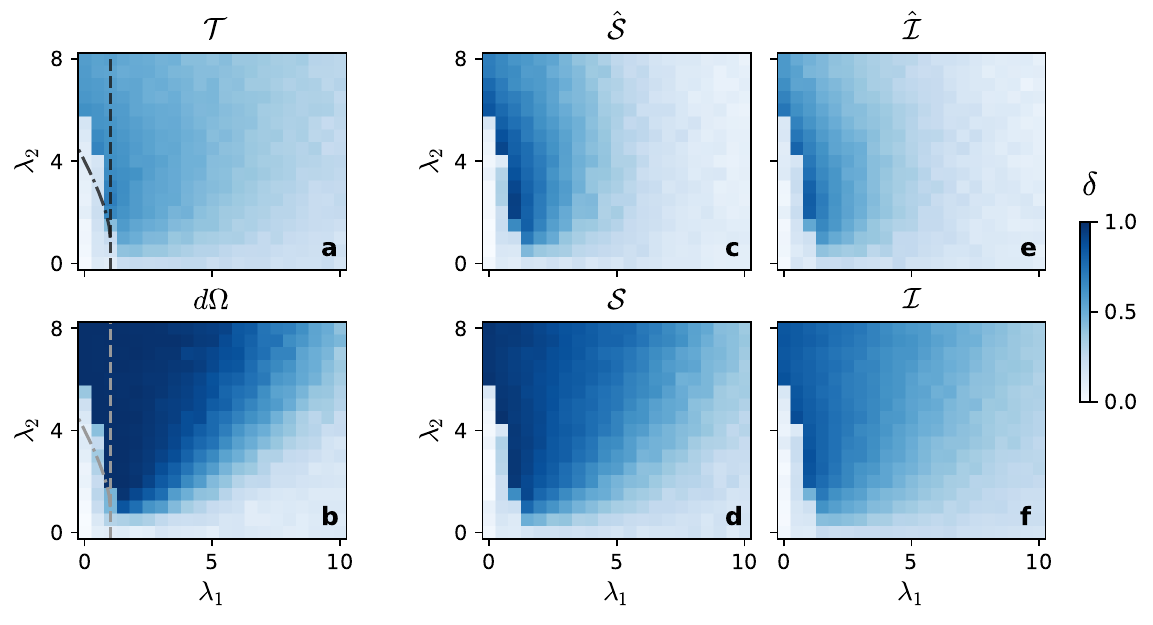}
    \caption{\textbf{Higher order behaviors in 2-simpleces.} Statistical distance $\delta$ between 3-clique and 2-simplex distributions for different lower and higher order information measures: \textbf{(a)} sum of pairwise transfer entropies $\mathcal{T}(\vb{X})$; \textbf{(b)} average dynamical O-information $d\Omega(\vb{X})$; \textbf{(c-d)} unconditioned $\hat{\mathcal{S}}$ and conditioned $\mathcal{S}$ average synergy; \textbf{(e-f)} unconditioned $\hat{\mathcal{I}}$ and conditioned $\mathcal{I}$ average mutual information.
    Phase space is composed of the pairwise and 2-simplex rescaled infectivities $(\lambda_1,\lambda_2)$ in the simplicial SIS model.
    Lines in panels \textbf{a} and \textbf{b} correspond to the mean-field critical infectivity due to pairwise (\textbf{dashed}) and higher-order (\textbf{dash-dotted}) mechanisms.
    }
    \label{fig:measures-tri}
\end{figure*}

Figure \ref{fig:measures-tri} shows how the pairwise $\lambda_1$ and higher-order $\lambda_2$ infection parameters in the simplicial SIS model affect the distance $\delta$ for the six measures of HOB defined previously.
We use a random simplicial complex of maximum order $L=2$, with generalized average degrees $\langle k_1\rangle=20$ and $\langle k_2 \rangle=6$.
The distance $\delta$, which quantifies the difference between HOB for cliques and simplices, depends non-trivially on $(\lambda_1, \lambda_2)$, in line with previous works \cite{robiglioSynergisticSignaturesGroup2025}.
When $\delta \approx 0$, we are unable to distinguish between pairwise and higher-order mechanisms, either because the measure is inappropriate or because there is little information in the observed dynamics to begin with. 
The practical implication is that in this case our ability to infer the presence of higher-order mechanisms from state dynamics alone, such as when working with empirical data, is limited. 

The regions of low ($\delta\approx0$) and high ($\delta > 0$) distinguishability correspond to regions where the pairwise or higher-order mechanism (respectively) is driving the dynamics and is therefore the dominant factor determining the system behavior. 
We see a strong response (large $\delta$) for all measures in two regions. The first is below the critical infectivity for pairwise interactions ($\lambda_1 < \lambda_1^*=1$, indicated by a dashed line in Fig. \ref{fig:measures-tri}\textbf{a-b}), but above that for the HO interactions ($\lambda_1 > 2\sqrt{\lambda_2}-\lambda_2$ \cite{iacopiniSimplicialModelsSocial2019}, indicated by a dash-dotted line in Fig. \ref{fig:measures-tri}\textbf{a-b}). 
In this region, simplices are entirely responsible for sustaining the infection, since below $\lambda_1^*$ it would die out if only pairwise mechanisms were at play. 
(For the full phase-space of the prevalence $\rho$ see Supplementary Figure 2.) 
The second region with large $\delta$ occurs for $\lambda > \lambda_1^*$ whenever the higher-order infectivity $\lambda_2$ is sufficiently larger than $\lambda_1$.
As we increase $\lambda_1$, pairwise mechanisms account for an increasing number of infection events, and thus only sufficiently strong higher-order mechanisms will create enough dependence on the group level to appear as information in the joint state of nodes. 
This is captured by all higher-order measures except the non-conditioned versions of the average synergy $\hat{\mathcal{S}}$ (Fig. \ref{fig:measures-tri}\textbf{c}) and average mutual information $\hat{\mathcal{I}}$ (Fig. \ref{fig:measures-tri}\textbf{e}).
For very large $\lambda_1$, the response for simpleces approaches that of cliques and thus $\delta\to0$. 
In this region, a larger proportion of infection events occur due to pairwise interactions, so HOMs have only a marginal effect on the dynamics, and the joint node state loses the extra information it provided when pairwise infection events were rare. 
In the subcritical regime ($\lambda_1<\lambda_1^*$), the transition from low to high distinguishability due to higher-order interactions follows the first-order transition between extinction and endemicity (i.e., non-zero prevalence) in the simplicial SIS model \cite{iacopiniSimplicialModelsSocial2019}. 
In the supercritical regime ($\lambda_1>\lambda_1^*$), the more gradual transition to distinguishability is not accompanied by a change in the prevalence $\rho$, so it is purely an effect of the competition between pairwise and higher-order mechanisms. 

Let us compare the response of different measures. 
While the purely lower-order metric $\mathcal{T}$ (Fig. \ref{fig:measures-tri}\textbf{a}) captures the essential characteristics described above, $d\Omega$ (Fig. \ref{fig:measures-tri}\textbf{b}) produces the clearest difference between cliques and simpleces throughout the phase space.
Interestingly, also the average mutual information $\mathcal{I}$ (Fig. \ref{fig:measures-tri}\textbf{f}) and especially the average synergy $\mathcal{S}$ (Fig. \ref{fig:measures-tri}\textbf{d}) produce a signal that is almost as strong as that of $d\Omega$. 
Since $d\Omega$ is equivalent to a balance between synergy and redundancy (see Eq. (9) in Section ``Dynamical O-information" of SM), synergy plays a more important role in distinguishing between lower- and higher-order mechanisms (see Supplementary Figure 1). 
Crucially, the effectiveness of both synergy and mutual information relies on conditioning: specifically, it relies on isolating the dynamic influence by conditioning on the target variable's history (e.g., conditioning on the target being susceptible in the SIS model).

Note that the average synergy $\mathcal{S}$, and especially the average mutual information $\mathcal{I}$, are much simpler measures than the dynamical O-information $d\Omega$. 
When thinking about these measures as a combination of partial information atoms (see Section ``Multivariate information and its decomposition" in SM), the synergy $S$ has a much simpler interpretation as the minimum amount of information contained in the state of all $n$ variables that is not contained in any subset of $n-1$ variables. 
As we increase the number of variables, $S$ continues to represent a single atom, while $d\Omega$ is a complex combination of atoms. 
At higher orders, it is easier to reason about $d\Omega$ as the difference between total correlation and dual total correlation \cite{rosasQuantifying2019}, which is still less intuitive than the definition of synergy. 
Lastly, if we are only interested in distinguishing between cliques and simpleces (e.g., for an inference problem), rather than analyzing the decomposition into different types of information, mutual information $\mathcal{I}$ appears as a decent and much less computationally intensive option 
(scaling as $n$, while $d\Omega(\vb{X})$ and $\mathcal{S}(\vb{X})$ scale as $n^2$). See Supplementary Figure 3 for the HOBs of a wider range of information-theoretic measures based on the Partial Information Decomposition.

\begin{figure}[tbp]
    \centering
    \includegraphics[width=\linewidth]{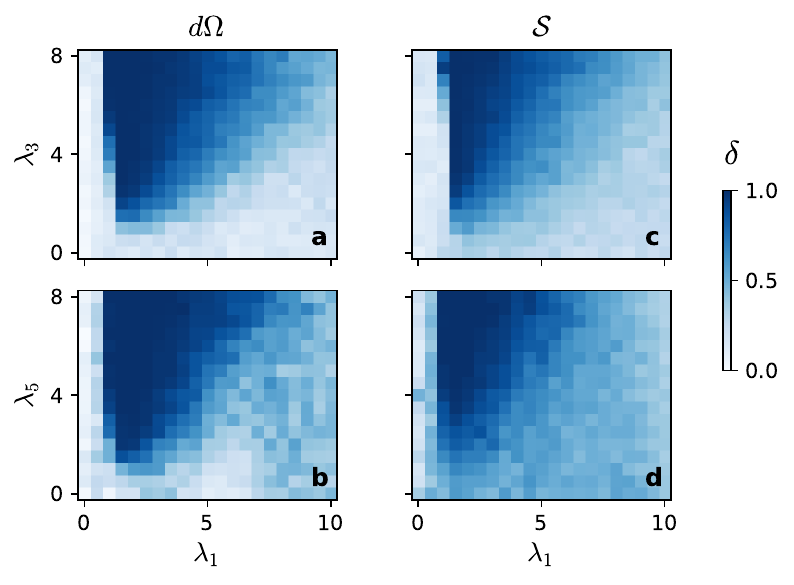}
    \caption{
    \textbf{Behaviors at higher orders.}
    Statistical distance $\delta$ for the average dynamical O-information $d\Omega$ (left column, \textbf{a-b}) and average synergy $\mathcal{S}$ (right column, \textbf{c-d}) distributions of $3$-hyperedges (top row, \textbf{a} and \textbf{c}) and $5$-hyperedges (bottom row, \textbf{b} and \textbf{d}) relative to their respective $(\ell+1)$-cliques. Distributions are based on separate simulations on random simplicial complexes with maximum order $L\in\{3,5\}$ and all mechanisms of intermediate order disabled, i.e. $\lambda_\ell=0$ for $2 \leq \ell < L$, effectively leaving only pairwise and order $L$ edges.
    }
    \label{fig:measures-higher-orders}
\end{figure}


Therefore, simpler measures such as average synergy $\mathcal{S}$ or even average mutual information $\mathcal{I}$, if conditioned to the history of the system, provide a sufficiently strong signal to distinguish the emergence of HOBs.
These observations are supported by Figure \ref{fig:measures-higher-orders}, showing $d\Omega(\vb{X})$ and $\mathcal{S}(\vb{X})$ for an $L=3$ and $L=5$ simplicial complex. 
We set the generalized average degrees to $\langle k_1\rangle=20$ and $\langle k_L \rangle=6$. 
We impose that only pairwise edges and the largest faces in the simplex are active (only $\lambda_1$ and $\lambda_L$ are nonzero). 
As before, we focus on how the difference $\delta$ between HOBs for $(L+1)$-cliques with and without $L$-simplices depends on the parameters $\lambda_1$ and $\lambda_L$, controlling the strength of pairwise and higher-order interactions, respectively. 
Note that applying HOB measures to a larger set $\vb{X}\equiv \vb{X}^n$ of $n=L+1$ variables introduces computational challenges (see Section ``Entropy estimation from data" in SM for discussion).

 Figure \ref{fig:measures-higher-orders} shows phase spaces similar to the $L=2$ case.
We can see regions with a large difference between the behaviors of pairwise cliques and simpleces ($\delta>0$) and regions where it is difficult to distinguish between the two ($\delta\approx0$). 
We also see the two types of transitions between these regions, for small and large pairwise infectivity $\lambda_1$. 
However, in this case, the subcritical transition for $\lambda_1 < 1$ does not occur when $\lambda_1\approx0$ and depends less strongly on the higher-order infectivity $\lambda_L$.

The average synergy $\mathcal{S}$ captures only a fraction of the mutual information between variables, unlike the more complex $d\Omega$. \
While this simplicity makes $\mathcal{S}$ less sensitive to HOBs near the $\lambda_1 = 1$ transition, it performs remarkably well elsewhere. 
Across the rest of the phase space, $\mathcal{S}$ distinguishes between cliques and simplices with the same precision as $d\Omega$, maintaining this parity at least up to order $L=5$.


\subsection{Cross-order induced behaviors}
\label{ssec:res-mixed}

In the previous section, we focused on measuring the behaviors at the maximum order $L$ of the simplicial complex, i.e., disabling all mechanisms except those at first and $L$-th order. 
We now look at a more realistic scenario, in which mechanisms of different orders are active at the same time, and we measure their effects on HOBs at all orders. 
We focus on random simplicial complexes of maximum order $L=3$ (tetrahedrons), which means we have three parameters: one lower-order parameter $\lambda_1$, and the higher-order parameters $\lambda_2$ and $\lambda_3$, controlling the strength of 2-simplices and 3-simplices, respectively. 
The networks have a size $N=200$ and average pairwise degree $\langle k_1 \rangle=20$ as before, and average generalized degrees set to $\langle k_2 \rangle=6$ and $\langle k_3 \rangle=2$. See Section ``Generating random higher-order networks" in SM for details on network generation. 
Recall that, in simplicial complexes, all simplices of intermediate orders (2-simplices, in this case) are located within structures of larger order.
We can now probe how the mechanisms at one order affect the behaviors at another order, by controlling the HOMs ($\lambda_2$ and $\lambda_3$).

\begin{figure}[tbp]
    \centering
    \includegraphics[width=\linewidth]{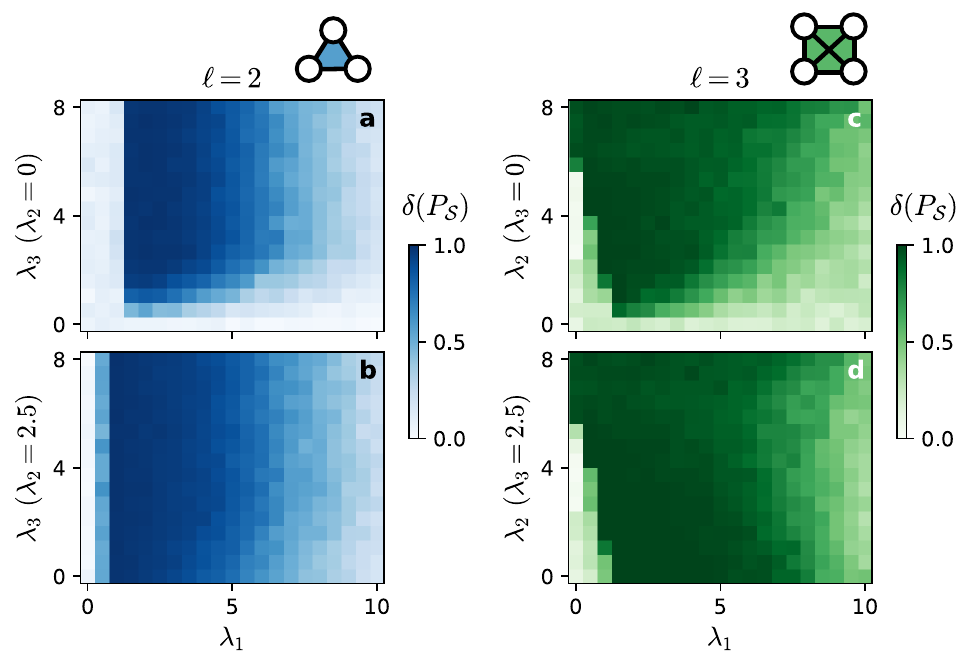}
    \caption{
    \textbf{Higher-order behaviors in mixed-order scenario.}
    Statistical distance $\delta$ between clique and simplex distributions of average synergy $\mathcal{S}$ for 2-simpleces (left, \textbf{a-b}) and 3-simpleces (right, \textbf{c-d}) depending on the strength of the other's mechanism in a random simplicial complex with maximum order $L=3$. Panels \textbf{a} and \textbf{c} show the HOBs at order $\ell$ when the $\ell$-th order mechanisms are disabled (equivalent to removing $\ell$-hyperedges). Panels \textbf{b} and \textbf{d} show the HOBs when both mechanisms are active.
    }
    \label{fig:mixed-syn}
\end{figure}

Figure \ref{fig:mixed-syn} shows how HOBs at order $\ell=2$ depend on $\lambda_3$, the strength of the order $\ell=3$ mechanism (panels \textbf{a} and \textbf{b}); and vice-versa, how HOBs at order $\ell=3$ depend on $\lambda_2$ (panels \textbf{c} and \textbf{d}).
We show the HOBs at order $\ell$ in two different cases: with $\ell$-th order mechanisms either disabled ($\lambda_\ell=0$) or active ($\lambda_\ell=2.5$). 
As before, we show the difference $\delta$ between the HOBs at $\ell$-simplices and $(\ell+1)$-cliques. 
We constrain the analysis to the average synergy $\mathcal{S}$, since we have seen it produces a similar signal to the dynamical O-information $d\Omega$ while being a simpler measure, especially for higher orders.


One can see strong HOBs at order $\ell$ even when the mechanism at that order is disabled ($\lambda_\ell=0$), in panels \textbf{a} and \textbf{c} of Figure \ref{fig:mixed-syn}.
A priori, we would expect that the HOBs observed by applying an information-theoretic measure will only be sensitive to mechanisms at the same order. 
Instead, we see that HOMs acting on one order can induce HOBs at other orders, affecting the dynamic response of the system. 
Therefore, in some contexts, a pure higher-order dynamics (HOM) is indistinguishable from induced HOBs at lower or higher orders.
Let us first focus on HOBs at order $\ell=2$. 
When $\lambda_2=0$, i.e., no activity due to 2-simplices (panel \textbf{a}), for sufficiently large pairwise infectivity $\lambda_1$, we observe a large $\delta(P_\mathcal{S})$, due to 2-simplices having more synergy than 3-cliques. 
As expected, the difference $\delta$ increases with the strength of the mechanisms at 3-simplices $\lambda_3$. 
A similar observation can be made for HOBs at order $\ell=3$ (panel \textbf{c}), with the difference that we can observe HOBs even for $\lambda_1=0$, due to sufficient $\lambda_2$.
We saw qualitatively the same phase space in Figure \ref{fig:measures-tri}, for HOBs at order $\ell=2$, due to $\lambda_2$.
We obtain similar results by using the sum of transfer entropies $\mathcal{T}$, the dynamical O-information $d\Omega$, and the average mutual information $\mathcal{I}$ (see Supplementary Figures 4 and 5).

From the functional point of view, disabling a mechanism $\lambda_\ell$ is equivalent to removing the respective $\ell$-faces from the network (while keeping the other orders intact), so observing HOB at order $\ell$ is a signature of induced behaviors between orders. 
On one hand, this detaches the measure of HOBs from the presence of mechanisms at the same order, which might not be practical in applications such as network inference. 
On the other hand, this sensitivity makes these measures a particularly useful tool for studing how higher-order behaviors can emerge, whether they come from higher-order or pairwise mechanisms. 
Later on in this section, we test whether the overlap and nestedness between edges can account for this effect.


When we activate the 2-order mechanisms, i.e., $\lambda_2=2.5$ (panel \textbf{b}), two things change with respect to the inactive case ($\lambda_2=0$) described above (panel \textbf{a}). 
First, the transition to distinguishability due to pairwise interactions occurs for a lower $\lambda_1$. 
As we saw in the previous experiments, this HOB transition coincides with the onset of HOM activity.
Since 2-order mechanisms require fewer simultaneously infected nodes than 3-order mechanisms, the minimum prevalence $\rho$ due to pairwise mechanisms necessary for HOMs to start triggering infections is reduced, and the onset of activity occurs for lower pairwise infectivity $\lambda_1$.
Second, the transition driven by 3-simplex infectivity ($\lambda_3$) nearly disappears, rendering the $\delta(P_\mathcal{S})$ phase space independent of $\lambda_3$. 
In contrast, for HOBs at order $\ell=3$ (panel \textbf{d}), the transition due to $\lambda_2$ is only lost in the supercritical regime (large $\lambda_1$); in the subcritical regime (small $\lambda_1$), the transition remains unaffected by the activation of $\lambda_3 = 2.5$. 
This asymmetry suggests that 2-hyperedges play a more fundamental role in generating HOBs across all orders. This observation aligns with existing literature, which indicates that higher-order terms generally contribute less to simplicial SIS and related dynamical models than their lower-order counterparts \cite{meloni2026reducibility,skardal2020higher,iacopiniSimplicialModelsSocial2019}.
To understand the cause of induced HOBs, in particular how they are related to the overlap and nestedness between mechanisms, we perform a series of experiments on random hypergraphs (see Section ``Random hypergraphs" in SM for details on network generation). 
To apply the same methodology as in previous experiments, we must first select the specific lower-order structures against which hyperedges will be compared to define the distance $\delta$. 
For simplicial complexes, it was intuitive that for any $\ell$-simplex an equivalent lower-order structure of $\ell+1$ nodes is simply an $(\ell+1)$-clique. 
However, hyperedges are not in general supported by pairwise cliques in a random hypergraph, so disabling the hyperedge leaves us with $\ell+1$ randomly chosen nodes that may or may not be connected by pairwise interactions or be part of other higher-order structures. 
Therefore, instead of using $(\ell+1)$-cliques as before, we use a uniform sample of $\ell+1$ nodes from all available nodes in the network as a baseline for comparing to HOBs in hyperedges. 
That is, we now define $\delta(P_M)$ as $\delta(P_M^r,P_M^s)$, where $P_M^r$ and $P_M^s$ are the measure $M$ distributions for random nodes and hyperedges, respectively.

By removing the structural correlations of simplicial complexes, nodes in hyperedges should exhibit the same HOBs as random groups of nodes, when the respective hyperedges are disabled.
Indeed, in a random hypergraph with order $\ell\in\{2,3\}$ HOMs, we see a low $\delta$ and no induced HOBs (see Supplementary Figure 6).
To see why, consider the two main correlations missing in a random hypergraph compared to a random simplicial complex, both due to the inclusion criterion of simpleces: the presence of pairwise cliques under each hyperedge and the nestedness between hyperedges of different orders. 
By nestedness, we refer to the complete inclusion of an interaction of some order within an interaction of another order, or in general of a group of nodes within another.
We conjecture that the absence of pairwise cliques under HOMs does not explain the lack of induced HOBs in the random hypergraph, since any HOBs due to pairwise interactions are discounted by taking the difference $\delta(P_\mathcal{S}^c,P_\mathcal{S}^s)$ between cliques and simpleces. Instead, it is the nestedness that creates the apparently induced HOBs. More accurately, we observe induced HOBs because we apply our measures to groups of nodes that are strict subsets or supersets of some higher-order mechanism, whether or not those nodes themselves form a mechanism.


We test this hypothesis by explicitly studying HOBs for groups of nodes in the neighbourhood of higher-order mechanisms (hyperedges), which we call ``nested groups". 
We construct a nested group of order $\ell_B$ by starting from the nodes in a mechanism of order $\ell_M$ and either removing random nodes (if $\ell_B<\ell_M$) or recursively adding a random pairwise neighbour of the nodes in the current set (if $\ell_B>\ell_M$), until we get a set of the desired size. 
As a result, each nested group is either completely contained within a hyperedge 
or contains a full hyperedge within it. 
For example, consider a mechanism of order $\ell_M=3$. A nested group of order $\ell_B=2$ is a set of three nodes out of the four that make up the mechanism; at order $\ell_B=\ell_M=3$, the nested group is the nodes in the mechanism itself; and at order $\ell_B=4$, we take a nested group by joining the set of nodes in the mechanism with one of their pairwise neighbours.
To properly test for induced HOBs, in the following experiments we ensure that the nested groups do not themselves form a hyperedge, unless $\ell_B=\ell_M$.
To get a set of nested groups for the statistical analysis, we take one group per mechansism (hyperedge) in the network, for each order $\ell_B$.

\begin{figure*}
    \centering
    \includegraphics[width=0.85\linewidth]{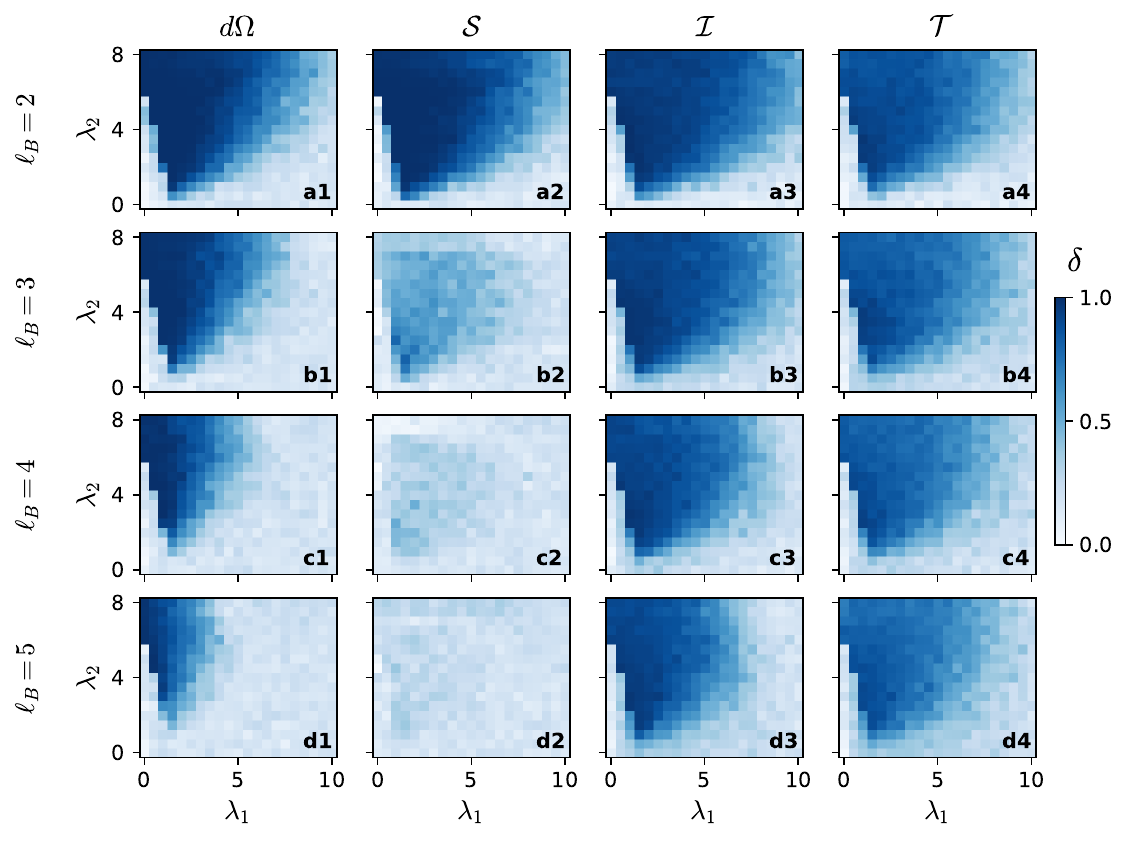}
    \caption{
    \textbf{Induced higher-order behaviours in nested hyperedges.}
    Statistical distance $\delta$ between HOB measure distributions for nested groups and random nodes, on a random hypergraph of maximum order $L=\ell_M=2$.
    Rows \{\textbf{a},\textbf{b},\textbf{c},\textbf{d}\} correspond to different orders $\ell_B\in\{2,3,4,5\}$ for the nested groups, respectively. In this case, row \textbf{a} corresponds to the actual order of the nested hyperedges, i.e. $\ell_B=\ell_M=2$.
    Columns correspond to different measures of HOB: average dynamical O-information $d\Omega$ (column \textbf{1}), average synergy $\mathcal{S}$ (column \textbf{2}), average mutual information $\mathcal{I}$ (column \textbf{3}) and sum of pairwise transfer entropies $\mathcal{T}$ (column \textbf{4}).
    }
    \label{fig:nested_3}
\end{figure*}

In Figure \ref{fig:nested_3} we show how different measures of HOB depend on the order $\ell_B$ of the nested group when there is a mechanism at order $\ell_M=2$. The setup involves a random hypergraph of $N=200$ nodes with average generalized degrees $\langle k_1\rangle=20$ and $\langle k_2\rangle=6$. 
For a given measure $M$, we show the statistical distance $\delta(P_M)\equiv\delta(P_M^r,P_M^n)$ between random and nested groups in the $(\lambda_1,\lambda_2)$ phase space of the simplicial SIS model. 

In general, almost all measures show induced HOBs at nested groups that do not correspond to the actual mechanism, i.e. we have $\delta\neq0$ when $\ell_B\neq\ell_M$. 
When comparing the response of different measures as we increase the order $\ell_B$ of the nested groups, we see clear differences in terms of distinguishability of nested from random groups.
The statistical distance $\delta$ for the average synergy $\mathcal{S}$ peaks at $\ell_B=\ell_M=2$ and decreases for $\ell_B>\ell_M$, as we can see in column \textbf{2} of Figure \ref{fig:nested_3}. This makes the synergy a confident indicator of the exact order of the underlying mechanism.
In contrast, the average mutual information $\mathcal{I}$ (column \textbf{3}) and sum of pairwise transfer entropies $\mathcal{T}$ (column \textbf{4}) show an almost constant distance $\delta$ as we take larger sets of variables (increase $\ell_B$). This implies we are able to distinguish groups that contain higher-order mechanisms but not their exact order.
The average dynamical O-information $d\Omega$ (column \textbf{1}) shows a balance between these two extremes, with a $\delta$ peaking at $\ell_B=\ell_M=2$. This is the case for all parameter values except close to the phase transition due to HOMs, where $d\Omega$ shows high $\delta$ for all orders of nestedness $\ell_B$. At higher orders, the O-information can be represented as a sum of information quantities at all lower orders within the group of source variables, making it sensitive to synergies even when they come from only subsets of the whole.
In Supplementary Figure 7 we show that these results are consistent for mechanisms at order $\ell_M=3$, although in that case the average synergy decreases less sharply whenever $\ell_B<\ell_M$.
There are two main takeaways from these results, if one is trying to infer the presence of HOMs by measuring HOBs.
First, if the order of mechanisms is unknown, the most reliable signal is achieved using the average synergy $\mathcal{S}$ due to its sensitivity to the actual order of mechanisms (see column \textbf{2} of Figure \ref{fig:nested_3}). Note that $\mathcal{S}$ is the only measure considered here that is also sensitive to higher-order mechanisms in its expected magnitude, not only the statistical distance $\delta$ (see Supplementary Figure 8).
Secondly, if the order of the target mechanisms is known and one simply has to find exactly where they are (which nodes compose them), it doesn't matter which measure one uses, as they provide a similar level of distinguishability (see row \textbf{a} of Figure \ref{fig:nested_3}). In this case, the most parsimonious and computationally efficient choice would be to use a measure like the average mutual information $\mathcal{I}$. 
However, the applicability of this scenario is limited to cases where only a single order of HOMs is present, since otherwise one could not distinguish mechanisms of different orders and the average synergy $\mathcal{S}$ would still be a better choice. 


\section{Conclusions}

We conducted a number of experiments relevant to both the application of measures of higher-order behavior (HOBs) for detecting higher-order mechanisms (HOMs), and the understanding of the relationship between the two.
While the dynamical O-information $d\Omega$ \cite{rosasQuantifying2019} is highly sensitive to the presence of HOMs for the simplicial SIS model on random simplicial complexes \cite{robiglioSynergisticSignaturesGroup2025}, simpler measures like the average synergy $\mathcal{S}$ do similarly well in capturing this behavior, also for orders up to $L=5$.
This suggests that to detect HOMs it might be sufficient to apply conceptually and computationally simple measures of higher-order information, including simply the average mutual information $\mathcal{I}$. 
Importantly, one needs to isolate the dynamical dependence between source and target variables (inherent in the $d\Omega$), since without removing intrinsic target dynamics, the measures show qualitatively different dependence and are no better than pairwise-only measures at distinguishing between lower and higher order structures.

Furthermore, we showed how higher-order behaviors can be induced by higher-order mechanisms across orders of interaction, meaning that a higher-order behavior can be observed for a group of nodes without a corresponding mechanism. 
Notably, hyperedge overlap is not necessary for observing these induced HOBs in our setup. 
Instead, induced behaviors can be measured for any group of nodes that either contains or is contained in a HOM, even if that mechanism is uncorrelated with interactions at other orders.
We tested how the observed HOBs change when measured for groups of varying sizes in the vicinity of HOMs, and find that induced HOBs are common for all measures, but some, like the average synergy $\mathcal{S}$, only show a strong signal when measured at the order of the underlying mechanisms.
Our work helps to highlight the limitations inherent in using statistical signatures to infer hidden interaction rules, demonstrating that higher-order behaviors do not always provide a direct mapping to their underlying mechanisms. 
By clarifying these boundaries and showing that signatures can emerge at scales where no dynamical rule is present, we highlight the need for a careful approach to higher-order inference when performing structural and dynamical analyses of complex systems. 


As an early work in this direction, our study is inherently limited in its scope. 
We focus on a single dynamical process (simplicial SIS model) and a single network structure (ER-type random higher-order networks).
The methodology we apply is based on estimation of information-theoretic quantities, which inherently requires a large amount of data and has known biases \cite{paninskiEstimationEntropyMutual2003,gehlenBiasOInformationEstimation2024}. 
In our specific setting, while we show that HOBs can be induced in nested groups, we do not explicitly show whether they can also be observed in groups that are partially overlapping with HOMs, although our results on random hypergraphs suggest this (see Supplementary Figure 6).

The two main avenues of future work are assessing the generalizability of our results to different experimental settings and testing the methodology with empirical data.
On one hand, working with different idealized models for dynamics on networks allows to test the robustness of the findings and gain a better understanding of the relationship between HOBs, HOMs, and specific dynamical rules. For example, whether all types of interactions show induced HOBs and whether the most appropriate measure of HOB depends on the specific dynamical process. 
On the other hand, it would be insightful to apply this methodology in empirical settings (with limited or noisy observations), in order to assess how severe the inherent data quantity requirements are. This would be a key step towards applicability in network inference.
A different direction, that is currently unexplored, is the relationship between HOBs and the structural properties of the groups of nodes, such as pairwise connectedness, centrality within the network, individual degree or degree heterogeneity, and hyperedge overlap \cite{maliziaHyperedgeOverlapDrives2025,Zhang2023synch}. 

\section{acknowledgments}
S.M. acknowledges support from the project `CODE - Coupling Opinion Dynamics with Epidemics', funded under PNRR Mission 4 `Education and Research' - Component C2 - Investment 1.1 - Next Generation EU `Fund for National Research Program and Projects of Significant National Interest' PRIN 2022 PNRR, grant code P2022AKRZ9. Partial support has also been received from the Agencia Estatal de Investigaci\'on and Fondo Europeo de Desarrollo Regional (FEDER, UE) under project APASOS (PID2021-122256NB-C22) and the Mar\'ia de Maeztu program, project CEX2021-001164-M, funded by the  MCIN/AEI/10.13039/501100011033.
M. S. acknowledges support from Grants No. RYC2022-037932-I and CNS2023-144156 funded by MCIN/AEI/10.13039/501100011033 and
the European Union NextGenerationEU/PRTR.
\bibliography{refs}

\end{document}


\title{Supporting Information: \\ 
Cross-order induced behaviors in contagion dynamics on higher-order networks}

\author{Kaloyan Danovski}
\affiliation{Department of Engineering, Universitat Pompeu Fabra, 08018 Barcelona, Spain}

\author{Sandro Meloni}
\affiliation{Institute for Cross-Disciplinary Physics and Complex Systems (IFISC, CSIC-UIB), 07122 Palma de Mallorca, Spain}
\affiliation{Institute for Applied Mathematics ``Mauro Picone'' (IAC) CNR, Via dei Taurini 19, 00185 - Rome, Italy}
\affiliation{``Enrico Fermi'' Research Center (CREF), Via Panisperna 89A, 00184 - Rome, Italy}

\author{Michele Starnini}
\email{Corresponding author: michele.starnini@upf.edu}
\affiliation{Department of Engineering, Universitat Pompeu Fabra, 08018 Barcelona, Spain}

\date{\today}

\maketitle

\section{Information-theoretic measures: Definition}
\label{ssec:method-info}


Here we describe the key measures of multivariate and higher-order information that we use in our analysis. These are built on standard measures in information theory like the entropy $H$ and mutual information $I$ \cite{coverElementsInformationTheory2006}. (All logarithms are base 2, and thus all measures have units of \textit{bits}.)

\subsection{Mutual information for dynamical processes}

We associate to each node $i$ in the network a dynamical random variable $X_i(t)$ following a distribution $P_i(x)$, representing the state of the node Susceptible/Infected $\mathcal{X}=\{0,1\}$.

Since we are considering dynamical models, we need to consider the directed information between the state of nodes with a certain time lag. For two variables $X$ and $Y$ this is captured by the \textit{time-delayed mutual information} (TDMI)
\begin{equation}\label{eq:tdmi}
    \hat{I}(X\to Y) \equiv I(Y^{(t)};X^{(t-1)}).
\end{equation}
Over the work we constrain the definition to interactions within single time step, but the extension to processes with longer memory is straightforward (although more computationally challenging, see Section ``Entropy estimation from data"). Note that the TDMI $\hat{I}$ is not necessarily symmetric, since in general $I(Y^{(t)},X^{(t-1)})\neq I(X^{(t)},Y^{(t-1)})$. 
In this context, we will refer to $Y$ as the \textit{target} variable, and $X$ as the \textit{source}.


However, the TDMI \eqref{eq:tdmi} is not a pure measure of interaction between variables, since it does not exclude the effects of intrinsic target dynamics and common drivers (confounder variables) when quantifying information between variables. To do this, we condition on the history of the target obtaining the well-known measure of \textit{transfer entropy} $T$ \cite{schreiberMeasuringInformationTransfer2000}
\begin{equation}\label{eq:te}
    I(X\to Y) \equiv I(Y^{(t)};X^{(t-1)}|Y^{(t-1)}) \equiv T(X\to Y).
\end{equation}


\subsection{Multivariate information and its decomposition}


In order to study higher-order behaviors, we need to analyze the interactions between three or more variables at the same time. The most straightforward extension to multivariate is to consider the joint state of a set of variables as a single unit for the pairwise mutual information $I$. For example, in a dynamical setting the amount of information that a set of $n$ sources $\vb{X}$ carries about a target $Y$ is quantified by
\begin{equation}\label{eq:mi_set}
    I(\vb{X}\to Y) \equiv I(Y^{(t)};\vb{X}^{(t-1)}|Y^{(t-1)}) \equiv T(\vb{X} \to Y),
\end{equation}
where $\vb{X}^{(t-1)}=\{X_i^{(t-1)}\}_{i=1..n}$. 
In this formulation, we treat a set of variables as a single unit, so we are still describing only pairwise influence. Fundamentally, this measure is unable to distinguish between different types of dependence \cite{jamesMultivariateDependenceShannon2017}. For example, it gives the same value both whenever the state of one variable in $\vb{X}$ completely determines $Y$ and whenever all sources in $\vb{X}$ are required to determine $Y$.

Thus, to quanify HOBs, we need to consider the multiple types of dependence that are aggregated into the multivariate mutual information. We expect in the presence of HOMs this information to be contained mainly in the joint state of all variables, as otherwise the dependence could be reduced to lower-level interactions. 
The information that is contained in the joint state of variables but not in the marginals is commonly called \textit{synergistic information} or simply \textit{synergy}. In contrast, information that can be obtained from any of the available sources is commonly called \textit{redundancy}. 
To distinguish between redundant and synergic interactions we employ Partial Information Decomposition (PID)  \cite{williamsNonnegativeDecompositionMultivariate2010}. 
For a target variable $Y$ and a set of $n$ sources $\vb{X}=\{X_i\}_{i=1..n}$, the PID assigns an information atom to each of a set of possible groups of source variables $\vb{X}_i$. These atomic quantities can be arranged on a lattice, where each node in the lattice represents the information available in a group of variables that is not available in a lower-level group. 
These give natural measures of synergy and redundancy, respectively, which are defined in the general case as
\begin{equation}
    S(Y;\vb{X})=I(Y;\vb{X})-\max_i{I(Y;\vb{X}_{-i})} \nonumber,
\end{equation}
\begin{equation}
    R(Y;\vb{X})=\min_i{I(Y;X_i)} \nonumber,
\end{equation}
where $\vb X_{-i}$ represents all the variables in $\vb X$ except for $i$, i.e. $\vb X_{-i}=\vb X \setminus \{X_i\}$.

These definitions fit well with the our intuitive understanding of synergy and redundancy. The measure $S$ gives us the information contained only in the joint state of all variables, and thus unrecoverable in any lower-level set. $R$ instead tells us the amount of information we could get by looking at any source variable, and thus the amount of redundant information at the level of pairwise dependence.

The fundamental insight of the PID framework is that the total amount of information $I(Y;\vb{X})$ 
that the sources $\vb{X}$ provide about the target $Y$ 
can be decomposed into a sum of positive, non-overlapping information quantities, two of which are the synergy $S$ and redundancy $R$ just defined. For two source variables, i.e. $\vb{X}=\{X_1,X_2\}$, the mutual information can be represented as
$$I(Y;\vb{X})=R(Y;\vb{X}) + U(Y;X_1) + U(Y;X_2) + S(Y;\vb{X}),$$
where $U(Y;X_i)$ is the \textit{unique} information that the source $X_i$ provides about the target $Y$.

To put these measures in the context of dynamical processes, we will introduce time delays between the dynamical variables and a conditioning on the target history. We can thus define the following quantities:
\begin{align}
    \hat{R}(Y;\vb{X}) &= \min_i{I(X_i\to Y)} \label{eq:red_nohist} \\
          R(Y;\vb{X}) &= \min_i{T(X_i\to Y)} \label{eq:red}
\end{align}
\begin{align}
    \hat{S}(Y;\vb{X}) &= I(\vb{X}\to Y)-\max_i{I(\vb{X}_{-i}\to Y)} \label{eq:syn_nohist} \\
          S(Y;\vb{X}) &= T(\vb{X}\to Y)-\max_i{T(\vb{X}_{-i}\to Y)} \label{eq:syn}
\end{align}

Finally, to characterize HOBs within a group of nodes, we use the average synergy or redundancy over all nodes in the group as targets. For example, for a hyperedge of order $\ell$ (size $n=\ell+1$), we compute $n$ times $R$ and $S$ varying the target node and then take the average. 



\subsection{Dynamical O-information}

Instead of looking at the amount of synergistic information, it is natural to ask what is the balance between synergistic and redundant information, as this quantifies how reducible (or not) the system is to lower-level interactions. 

For this purpose, Rosas et al. \cite{rosasQuantifying2019} introduced the \textit{O-information} $\Omega$, which was later extended by Stramaglia et al. \cite{stramagliaQuantifyingDynamicalHighOrder2021} in the context of dynamical processes to the \textit{dynamical O-information} $d\Omega$. The $d\Omega$ measures the difference between group synergy and sub-group redundancies in the influence of a set of $n$ source variables $\vb X=\{X_i\}_{i=1..n}$ on a target variables $Y$. It is defined as
\begin{align}\label{eq:do-info}
    d\Omega(Y;\vb X) &= (1-n)I\qty(Y^{(t)};\vb{X}^{(t-1)}|Y^{(t-1)})+\sum_{i=1}^n I\qty(Y^{(t)};\vb{X}_{-i}^{(t)}|Y^{(t-1)}) \\
                     &= (1-n)T(\vb{X}\to Y)+\sum_{i=1}^n T(\vb{X}_{-i}\to Y), \nonumber
\end{align}
We see that $d\Omega$ checks the amount of information in the whole group against that provided by all sub-groups of size $n-1$. 

$d\Omega$ has a few useful properties \cite{rosasQuantifying2019}. Firstly, the number of terms needed to calculate $d\Omega$ scales linearly with the system size $n$. In general, $d\Omega>0$ indicates a predominance of redundant information, while $d\Omega<0$ indicates a predominance of synergistic information. 
In the case of three variables (two sources and one target), it can be shown that it is equal to the difference between $R$ and $S$, resulting in:
\begin{equation}\label{eq:r-s}
d\Omega(Y;\vb{X})=R(Y;\vb{X})-S(Y;\vb{X}).
\end{equation}

Again, since $d\Omega$ is a directed measure, to measure HOBs within a group we consider the average of $d\Omega$ over all nodes in the group as targets and the rest as sources. 



\section{Generating random higher-order networks}
\label{ssec:method-nwk}
Throughout the work we consider both random simplicial complexes and random hypergraphs with the aim of assesing how the inter-order overlap affects the detectability of HOBs.  

\subsection{Random hypergraphs}

The random hypergraph model we employ is one of the possible extensions of the Erdos-Renyi (ER) random graph model where, for each order $\ell$ we fix the number of edges $M_\ell$.
To sample from this ensemble, we take $M_\ell$ uniform samples of $\ell+1$ nodes without replacement for each order $\ell$, and connect them. To obtain a network with target generalized degrees $k_\ell^*$ we can set $M_\ell=Nk_\ell^*/(\ell+1)$ in the sampling process described above. A function implementing this procedure is available in the \texttt{hypergraphx} Python package \cite{lotito2023hypergraphx}.


\subsection{Random simplicial complexes}
In order to understand the role of inter-layer overlap in the measured HOBs, along with random hypergraphs we also use simplicial complexes because the overlap between edges (of both higher and lower order) is predictable.


To generate random simplicial complexes, we start with a random hypergraph $H=(V,E_H)$, as described in the previous section. Then, we construct a new edge set $E_K$ out of every hyperedge $e\in E_H$ and all possible edges of lower order between the nodes in $e$. For example, if we have a 3-hyperedge, we will add all possible 2-hyperedges (triangles), as well as all possible 1-hyperedges (pairwise links). The resulting hypergraph $(V,E_K)$ will satisfy the inclusion criterion and thus be a simplicial complex $K$.

Same as when we generate random hypergraphs, we are interested in controlling the generalized average degrees $\langle k_\ell\rangle$ in the simplicial complex. However, this is complicated by the procedure described above, since in general not all edges of lower order will be present and including them will increase $\langle k_\ell\rangle$. To go around this, we adjust the number of edges $M_\ell$ of the ensemble $H(N,M_1,...,M_L)$ that we use to generate the initial hypergraph in order to obtain the target degrees $k_\ell^*$ in the simplicial complex.

We can do this as follows. For an edge of order $\ell+1$, each newly added hyperedge of order $\ell$ will increase the average degree $\langle k_\ell\rangle$ by some amount depending on the probability that edges of order $\ell$ are already present in $H$. This probability is $p_\ell=M_\ell/T(\ell)$, where $T(\ell)=\prod_{j=0}^\ell(N-j)/(\ell+1)$ is the total number of possible edges of order $\ell$ in a network of $N$ nodes. Assuming $p_\ell=M_\ell/T(\ell)\ll1$, i.e. a sparse network, we obtain the equation
\begin{equation}\label{eq:kmean}
    \langle k_\ell\rangle \approx \frac{(\ell+1)M_\ell}{N} + (\ell+1)\langle k_{\ell+1}\rangle \qty(1-\frac{M_\ell}{T(\ell)}).
\end{equation}
By plugging in our targets $k_\ell^*$ and inverting the equation, we obtain an expression for the target $M_\ell$ that will result in $\langle k_\ell\rangle\approx k_\ell^*$. Since no edges of the maximum order $L$ are added to the hypergraph, we can calculate the adjusted number of nodes $M_\ell$ recursively, starting with $\langle k_L\rangle=k_L^*$ and $M_L=N\langle k_L\rangle/(L+1)$.

For example, suppose we want to generate a simplicial complex with hyperedges up to order $L=2$ (triangles). We would begin by fixing the top-level average degree to $k_2^*$, which gives us a total of $M_2=Nk_2^*/3$ triangles. Then, from Eq. \eqref{eq:kmean} the expected pairwise (order $L-1=1$) degree is
$$
\langle k_1\rangle \approx \frac{2M_1}{N} + 2\langle k_2\rangle \qty(1-\frac{2M_1}{N(N-1)}).
$$
Rearranging this equation as
$$
M_1 = N(N-1)\frac{\langle k_1\rangle - 2\langle k_2\rangle}{2(N-1)-4\langle k_2\rangle}
$$
we can calculate the number of pairwise edges $M_1$ that we need for a desired pairwise degree $\langle k_1\rangle\approx k_1^*$. We use $M_1$ and the previously fixed $M_2$ as parameters to sample from the $H(N,M_1,M_2)$ ensemble of hypergraphs, and then fill in the missing lower-order edges to obtain a simplicial complex.




\section{Numerical simulations of the Contagion Dynamics}
\label{ssec:method-dyn}

The method we apply for quantifying HOBs can be used with empirical data on dynamics. However, to evaluate its potential systematically we restrict our study to numerical simulations of dynamical models on networks. In particular, we focus on the Susceptible-Infected-Susceptible (SIS) model of contagion \cite{pastor-satorrasEpidemicProcessesComplex2015}, and its extension to higher-order interactions \cite{iacopiniSimplicialModelsSocial2019}.


\subsection{Simplicial SIS model}

The SIS model posits that each of $N$ individuals (nodes) in a population (network) of can be in one of two states: susceptible or infected. A susceptible node is infected by each of its infected neighbours (separately) at a pairwise \textit{infection rate} $\beta$. Additionally, for each hyperedge $e$ of order $\ell$ that the node is a part of, it is infected at an additional rate $\beta_\ell$ if all other nodes in $e$ are infected. An infected node recovers at a rate $\mu$ independently of its neighbours.

We can represent nodes as stochastic variables with states $0$ (susceptible) or $1$ (infected). The state of node $i$ at a time $t$ is denoted as $\sigma_i(t)\in\mathcal{X}=\{0,1\}$, and the overall microstate of a system of $N$ nodes is the vector $\sigma(t)=[\sigma_1(t),\sigma_2(t),...,\sigma_N(t)]\in\mathcal{X}^N$. The macroscopic order parameter is the prevalence $\rho(t)=\sum_i\sigma_i(t)/N$. 

For the case of pairwise (order 1) interactions only, the mean-field approximation with homogeneous average degree $\langle k_1\rangle$ yields an \textit{epidemic threshold} for the infection rate $\beta_1^*=\mu/\langle k_1\rangle$. Below $\beta_1^*$ the disease will die out, while above it will persist in the population at a finite prevalence. Because of this, and in order to make it easier to compare between networks with different connectivity, we usually work with the \textit{rescaled infectivity} $\lambda_1=\beta_1\langle k_1\rangle/\mu$.
In the Simplicial SIS model by Iacopini et al. \cite{iacopiniSimplicialModelsSocial2019}, this reasoning is generalized to higher-order interactions by defining one rescaled infectivity parameter for each order as $\lambda_\ell=\langle k_\ell\rangle\beta_\ell/\mu$. The authors also show that show that there is a transition to endemicity due to second order interactions (2-hyperedges) at $\lambda_1^*=2\sqrt{\lambda_2}-\lambda_2$. These are the main control parameters of the model, which we interpret as controlling the strength of PM ($\ell=1$) and HOM ($\ell\geq2$).

\subsection{Numerical simulation procedure}
\label{ssec:method-dyn-sim}

There are many ways to obtain numerical approximations to the dynamical equations of the (simplicial) SIS model. In general one needs to be careful, since simulation schemes that result in otherwise equivalent macro-scale observables can show distinct micro-scale patterns, which in turn affect the estimation of probability-based measures like the ones described in Section ``Information-theoretic measures: definition". Recall that our information-based approach requires we reconstruct the time-lagged joint probability dsitrbutions over node states, which is much easier when dynamics are discrete and Markovian. Hence, we will focus on the simplest and most coarse approximation of the dynamics: a discrete-time, synchronous update agent-based simulation. 

The simulation proceeds as follows. At the initial time, each node starts in the infected state with a probability $\rho_0$. At each time $t$ afterwards, we use the current node states $\sigma(t)$ to obtain the following $\sigma(t+1)$. For each node $i$:
\begin{itemize}
    \item If $\sigma_i(t)=1$ (infected), we set $\sigma_i(t+1)=0$ with a probability $\mu$.
    \item If $\sigma_i(t)=0$ (susceptible), node $i$ has a probability to be infected from each of its interactions. For each edge $e$ of order $\ell$ incident on $i$ where all other nodes are infected, i.e. $\sigma_j(t)=1$, $\forall j\in e \text{ s.t. } j\neq i$, we have an independent probability $\beta_\ell$ that $\sigma_i(t+1)=1$.
\end{itemize}
This mechanism is analogous to the simplicial contagion model by Iacopini et al. \cite{iacopiniSimplicialModelsSocial2019}, except that the latter was defined only for simplicial complexes, and we simply assume that the independent higher-order infection rules apply in an analogous way in hypergraphs.

For all of the simulations presented in the main text, we fix the parameters $\mu=0.8$ and $\rho_0=0.3$, while varying the infection rates $\beta_\ell$ (by controlling $\lambda_\ell$).

\section{Information-theoretic measures: Measurement}
\label{ssec:method-data}


In this section, we summarize some of the main computational techniques and give some relevant definitions.

\subsection{Entropy estimation from data}

From the simulation procedure described in the previous section we obtain a series of vectors $\sigma(t)$ of dimension $N$, which we use to estimate the information measures defined in Section ``Information-theoretic measures: definition". 

The easiest and most agnostic approach to this is to estimate probability terms as $\hat p(x)=n_x/n_{tot}=\sum_t\sigma_X(t)/t_{\max}$, or the number of times the state $x$ was observed for variable $X$ in the data $\sigma_X$, divided over the total number of samples $t_{\max}$. Then, the simplest approach is to employ the maximum likelihood estimator (MLE) for the entropy $H$, which is defined (for a single binary variable $X$) as:
$$
\hat H_{MLE}=\sum_{x\in\mathcal{X}} \hat p(x) \log \hat p(x).
$$
This approach is straightforwardly extended to joint and conditional distributions, and thus to arbitrary entropy, mutual information, and other higher-order information measures.

For an overview of entropy and information estimation methods, see Ref \cite{paninskiEstimationEntropyMutual2003}. In additoin, we note that the bias in using the MLE approach to estimate the O-information was explored by Gehlen et al. \cite{gehlenBiasOInformationEstimation2024}, where they find that a naive O-information estimation is consistently biased toward synergy $\Omega<0$ with insufficient samples, especially for systems of independent variables (where we expect $\Omega=0$). The accuracy of the estimated entropy depends on the amount of observational data we have, which for empirical data tends to be scarce or expensive. The curse of dimensionality exacerbates this issue when increasing system size, since we need the convergence of $|\mathcal{X}|^n$ values for the joint distribution of $n$ variables over a shared state space $\mathcal{X}$. In our experiments we find that even for the simplest cases (3 variables) we require on the order of $10^4$ samples to clearly observe the statistical patterns.

\subsection{Distance between distributions}

We use an aggregate measure for the distance between distributions in order to quantify the difference between the higher-order behavior signals of different groups of nodes, for example pairwise cliques and simpleces. Following previous work \cite{robiglioSynergisticSignaturesGroup2025}, we use the \textit{statistical distance} $\delta$. For two discrete distributions $P_1,P_2$ over a common, discrete domain $\mathcal{Z}$, the \textit{statistical distance} $\delta$ is defined as
\begin{equation}\label{eq:statd}
    \delta(P_1,P_2)=\frac{1}{2}\sum_{z\in\mathcal{Z}} |P_1(z)-P_2(z)|.
\end{equation}
In practice, we need to discretize the joint support of the two distributions into $K$ bins, thus obtaining a discrete support set $\mathcal{Z}$ and then summing over the probability values in each bin.

Our code for the simplicial SIS simulations, estimation of information measures, and  data analysis, is available on GitHub \cite{induced-hob}.

\section{Additional results}

Below we present a number of supporting figures that have been referenced in the main text of the paper.



\begin{figure}
    \centering
    \includegraphics[width=\linewidth]{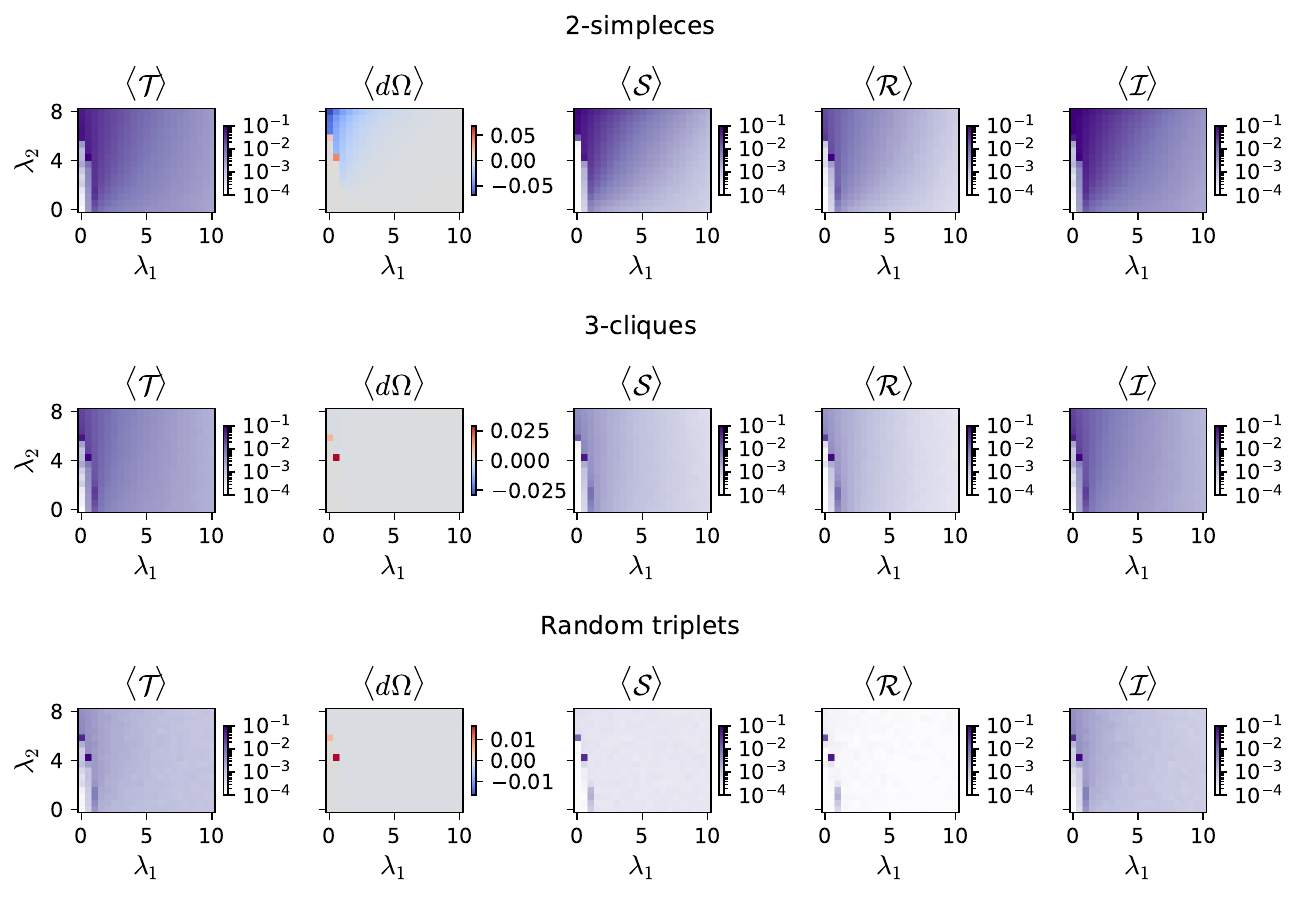}
    \caption{
    \textbf{Magnitude of HOB measures in 2-simpleces, 3-cliques, and random triplets.}
    Quantified as the mean value of the measure distributions for different groups (corresponding to rows). Columns correspond to different measures of HOB: sum of pairwise transfer entropies $\mathcal{T}$, average dynamical O-information $d\Omega$, average synergy $\mathcal{S}$, average redundancy $\mathcal{R}$, average mutual informaion $\mathcal{I}$.
    Phase space is composed of the rescaled infectivity parameters $(\lambda_1,\lambda_2)$ for the simplicial SIS model.
    }
    \label{fig_sm:measure_abs}
\end{figure}

\begin{figure}
    \centering
    \includegraphics[width=0.75\linewidth]{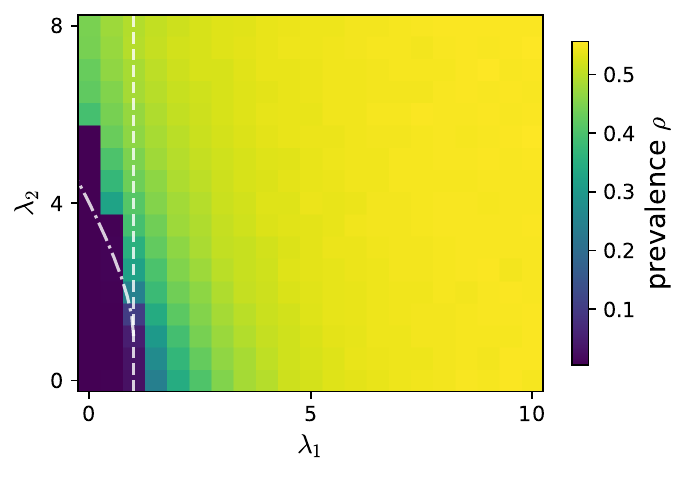}
    \caption{
    \textbf{Steady-state prevalence in the simplicial SIS model.}
    Results are shown in the phase space of the lower and higher order rescaled infectivity parameters $\lambda_1$ and $\lambda_2$. White lines show the values of $(\lambda_1,\lambda_2)$ at which we observe a critical transition from extinction to endemicity due to pairwise (\textbf{dashed} line) or higher-order (\textbf{dash-dotted} line) infectivity in the mean-field model, as described in 
    Ref. \cite{iacopiniSimplicialModelsSocial2019}.}
    \label{fig:prev}
\end{figure}

\begin{figure}
    \centering
    \includegraphics[width=0.8\linewidth]{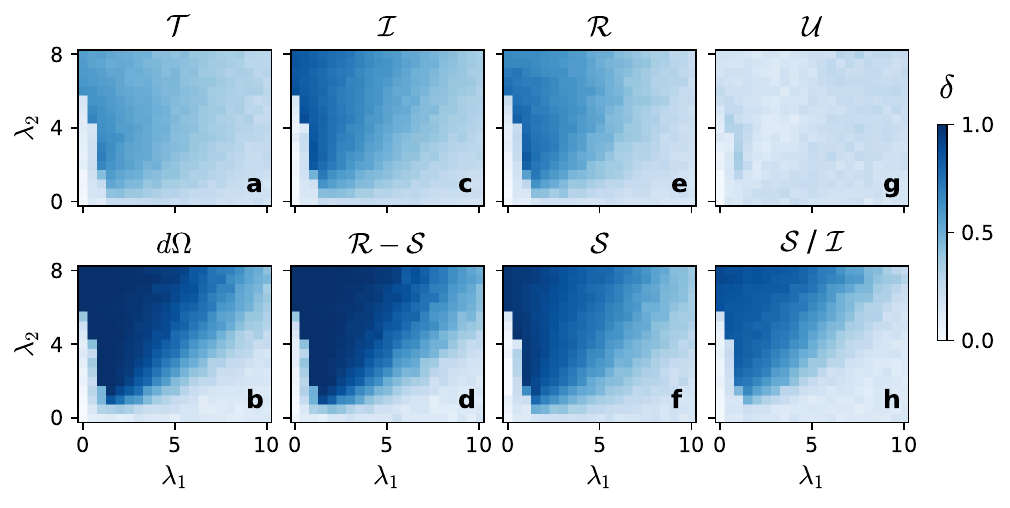}
    \caption{
    \textbf{Full range of measures showing higher-order behaviors on 2-simpleces.}
    Higher-order behaviors measured by the statistical distance $\delta$ between 3-clique and 2-simplex distributions for different information-theoretic measures: (\textbf{a}) sum of pairwise transfer entropies $\mathcal{T}$, (\textbf{b}) average dynamical O-information $d\Omega$, (\textbf{c}) average mutual informaion $\mathcal{I}$, (\textbf{d}) difference between average redundancy and average synergy $\mathcal{R}-\mathcal{S}$, (\textbf{e}) average redundancy $\mathcal{R}$, (\textbf{f}) average synergy $\mathcal{S}$, (\textbf{g}) average sum of unique informations $\mathcal{U}=\mathcal{I}-\mathcal{R}-\mathcal{S}$, (\textbf{h}) average synergy ratio $\mathcal{S}/\mathcal{I}$.
    Phase space is composed of the rescaled infectivity parameters $(\lambda_1,\lambda_2)$ for the simplicial SIS model.
    }
    \label{fig_sm:distances_all}
\end{figure}

\begin{figure}
    \centering
    \includegraphics[width=0.9\linewidth]{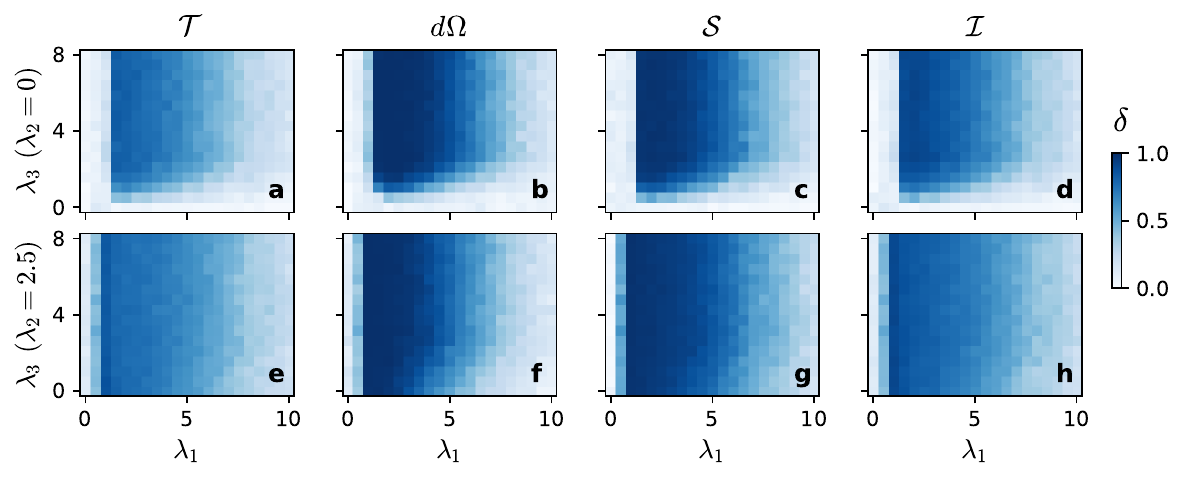}
    \caption{
    \textbf{Higher-order behaviors at order 2 induced by 3-mechanisms.}
    Statistical distance $\delta$ between 3-clique and 2-simplex distributions for different measures of HOB: \textbf{(a,e)} sum of pairwise transfer entropies $\mathcal{T}$, \textbf{(b,f)} average dynamical O-information $d\Omega$, \textbf{(c,g)} average synergy $\mathcal{S}$, \textbf{(d,h)} average mutual informaion $\mathcal{I}$.
    HOBs are shown as a function of the strength of the 3-mechanism $\lambda_3$ in two cases: \textbf{(a-d)} with 2-order mechanisms disabled ($\lambda_2=0$, equivalent to removing the 2-mechanisms) and \textbf{(e-h)} with mechanisms at both orders active ($\lambda_2=2.5$).
    }
    \label{fig_sm:mixed_all_2}
\end{figure}

\begin{figure}
    \centering
    \includegraphics[width=0.9\linewidth]{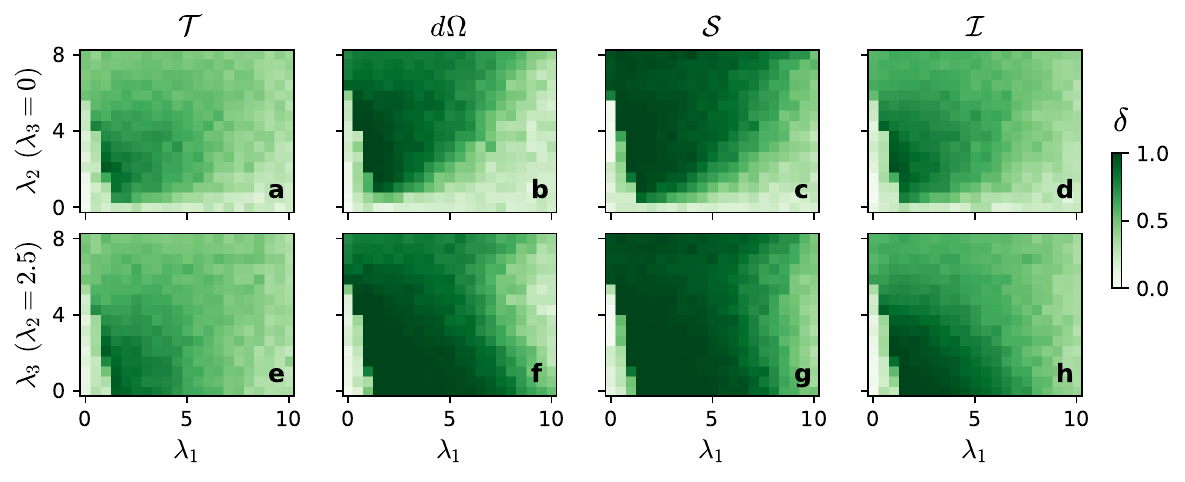}
    \caption{
    \textbf{Higher-order behaviors at order 3 induced by 2-mechanisms.}
    Statistical distance $\delta$ between 4-clique and 3-simplex distributions for different measures of HOB: \textbf{(a,e)} sum of pairwise transfer entropies $\mathcal{T}$, \textbf{(b,f)} average dynamical O-information $d\Omega$, \textbf{(c,g)} average synergy $\mathcal{S}$, \textbf{(d,h)} average mutual informaion $\mathcal{I}$.
    HOBs are shown as a function of the strength of the 2-mechanism $\lambda_2$ in two cases: \textbf{(a-d)} with 3-order mechanisms disabled ($\lambda_3=0$, equivalent to removing the 3-mechanisms) and \textbf{(e-h)} with mechanisms at both orders active ($\lambda_3=2.5$).
    }
    \label{fig_sm:mixed_all_3}
\end{figure}


\begin{figure}
    \centering
    \includegraphics[width=0.7\linewidth]{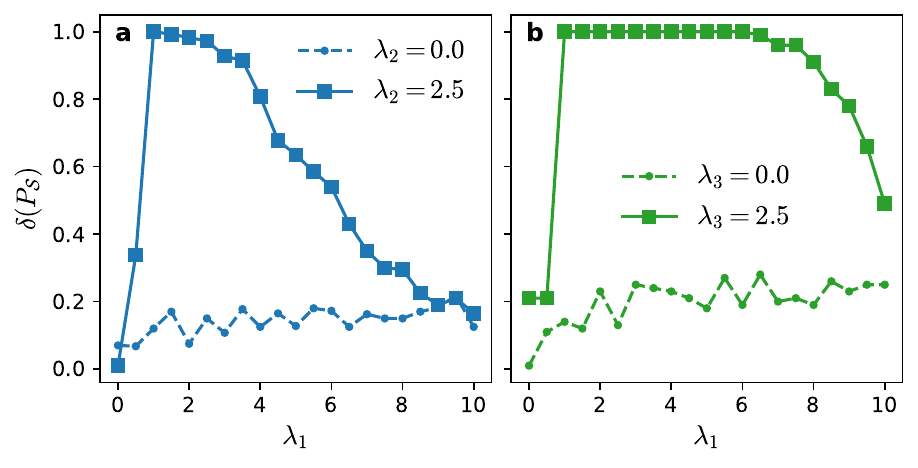}
    \caption{
    \textbf{No induced higher-order behaviours in independent hyperedges.}
    Statistical distance $\delta$ between synergy $\mathcal{S}$ distributions of $\ell$-hyperedges and groups of $\ell+1$ random nodes in random hypergraphs with maximum order $L=3$. 
    Results are for HOBs at order $\ell=2$ (panel \textbf{a}) and $\ell=3$ (panel \textbf{b}), colors are simply a visual aid. While modulating the strength of $\ell$-order mechanisms $\lambda_\ell$, we keep the mechanism at the other order fixed. Concretely, in panel \textbf{a} we change $\lambda_2$ for a fixed $\lambda_3=2.5$, while in panel \textbf{b} we fix $\lambda_2=2.5$ and change $\lambda_3$. Dashed lines show HOBs when $\ell$-order mechanisms are disabled ($\lambda_\ell=0$). Solid lines show HOBs when mechanisms at both orders are active.
    Dynamics on random hypergraphs are obtained analogously to the simplicial SIS model, with independent infection events for pairwise and higher-order structures.
    }
    \label{fig:mixed-hg-hyperedge}
\end{figure}

\begin{figure}
    \centering
    \includegraphics[width=0.8\linewidth]{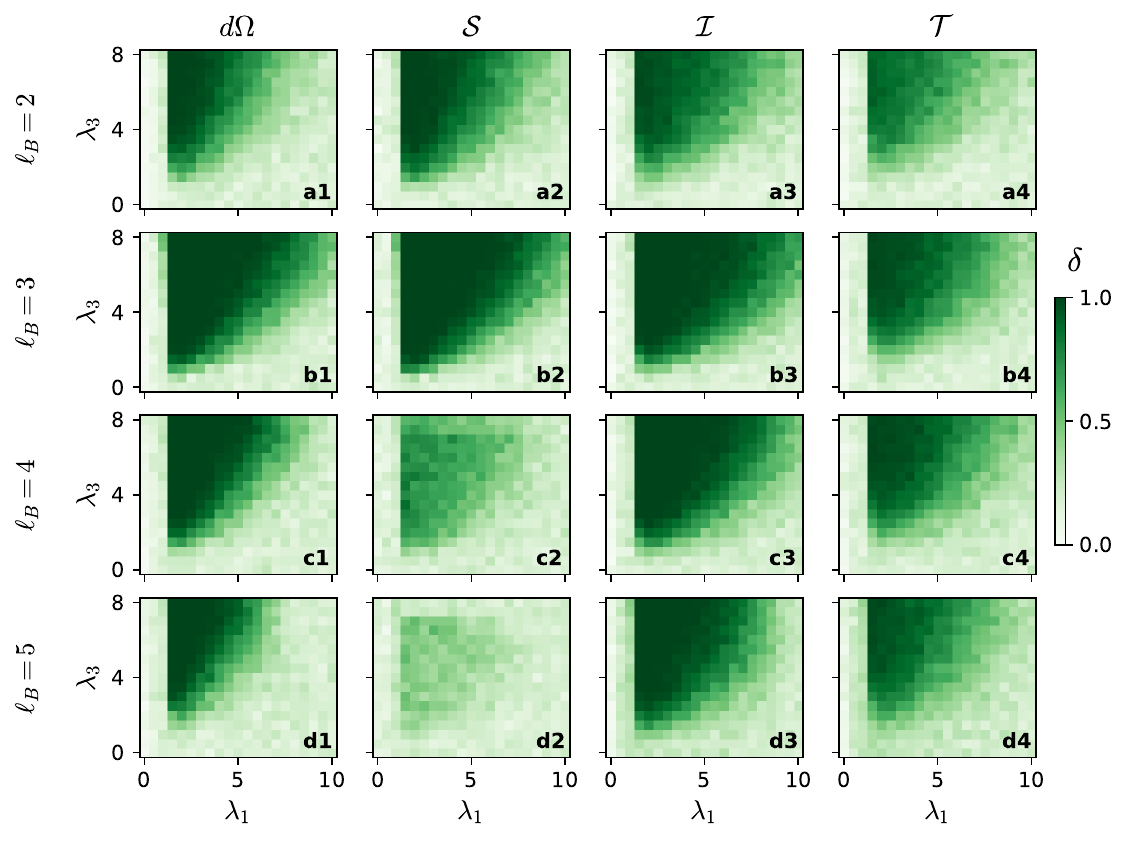}
    \caption{
    \textbf{Induced higher-order behaviours in nested hyperedges at order $\ell_M=3$.}
    Statistical distance $\delta$ between HOB measure distributions for nested groups and random nodes, on a random hypergraph of maximum order $L=\ell_M=3$ and no 2-hyperedges.
    Rows \{\textbf{a},\textbf{b},\textbf{c},\textbf{d}\} correspond to different orders $\ell_B\in\{2,3,4,5\}$ for the nested groups, respectively. In this case, row \textbf{b} corresponds to the actual order of the nested hyperedges, i.e. $\ell_B=\ell_M=3$.
    Columns correspond to different measures of HOB: average dynamical O-information $d\Omega$ (column \textbf{1}), average synergy $\mathcal{S}$ (column \textbf{2}), average mutual information $\mathcal{I}$ (column \textbf{3}) and sum of pairwise transfer entropies $\mathcal{T}$ (column \textbf{4}).
    The phase space consists of the rescaled infectivities $(\lambda_1,\lambda_3)$.
    Note the less pronounced decrease for the synergy $\mathcal{S}$ when $\ell_B<\ell_M$ (panel \textbf{a2}) compared to when $\ell_B>\ell_M$ (panels \textbf{c2} and \textbf{d2}) and the results for $\ell_M=2$ in the main text (Fig. 4).
    }
    \label{fig:nested_4}
\end{figure}

\begin{figure}
    \centering
    \includegraphics[width=0.99\linewidth]{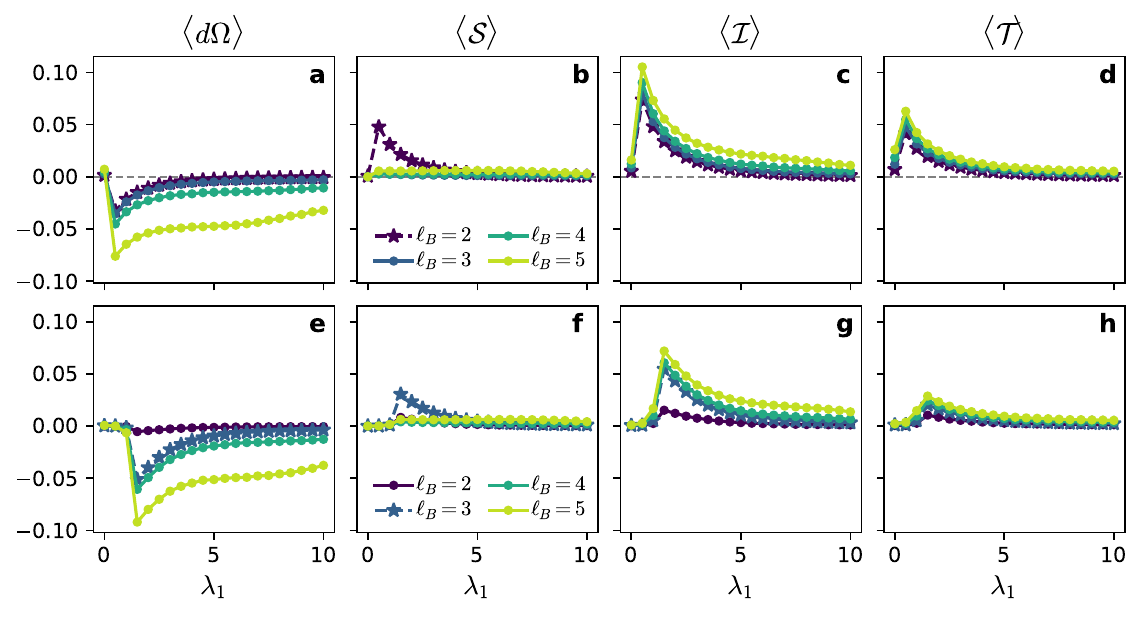}
    \caption{
    \textbf{Magnitude of HOB measures with nested hyperedges.}
    Mean value of measure distributions for different order of nested groups $\ell_B$ and different measures of HOB: \textbf{(a,e)} average dynamical O-information $d\Omega$, \textbf{(b,f)} average synergy $\mathcal{S}$, \textbf{(c,g)} average mutual informaion $\mathcal{I}$, \textbf{(d,h)} sum of pairwise transfer entropies $\mathcal{T}$.
    Results are for mechanisms of order $\ell_M=2$ (pabels \textbf{a-d}) and $\ell_M=3$ (panels \textbf{e-h}).
    Colors correspond to different values of $\ell_B$, with the case $\ell_B=\ell_M$ highlighted (\textbf{dashed stars}) in each scenario.
    Results are shown as a function of the pairwise infectivity $\lambda_1$ for a fixed higher-order infectivity $\lambda_{\ell_M}=5$.
    Note the pronounced increase in synergy when behaviors are measures at the actual order of the underlying mechanisms (stars in panels \textbf{b} and \textbf{f}).
    }
    \label{fig:nested_abs}
\end{figure}



\bibliography{refs}